\documentclass{osa-article}

\usepackage{amsfonts}
\usepackage{booktabs}
\usepackage{siunitx}
\usepackage{tabularx,ragged2e,booktabs} % for formatting tables
\usepackage{array}
\newcolumntype{L}{>{\centering\arraybackslash}m{3cm}}
\usepackage{chngcntr}

\journal{osajournal}
\articletype{Research Article}
\begin{document}

\title{Multi-functional integrated photonics in the mid-infrared with suspended AlGaAs on silicon}

\author{Jeff Chiles,\authormark{1,*} Nima Nader,\authormark{1} Eric J. Stanton,\authormark{1} Daniel Herman,\authormark{1,2} Galan Moody,\authormark{1} Jiangang Zhu,\authormark{3} J. Connor Skehan,\authormark{1,$\dagger$} Biswarup Guha,\authormark{4,5} Abijith Kowligy,\authormark{6} Juliet T. Gopinath,\authormark{2,3} Kartik Srinivasan,\authormark{4} Scott A. Diddams,\authormark{2,6} Ian Coddington,\authormark{1} Nathan R. Newbury,\authormark{1} Jeffrey M. Shainline,\authormark{1} Sae Woo Nam,\authormark{1} and Richard P. Mirin\authormark{1}}

%\author{Jeff Chiles,\authormark{1} Eric J. Stanton,\authormark{1} Nima Nader\authormark{1}}

\address{\authormark{1} NIST, Applied Physics Division, 325 Broadway, Boulder CO 80305, USA\\
\authormark{2} CU Boulder, Department of Physics, Boulder CO 80309, USA\\
\authormark{3} CU Boulder, Department of Electrical and Computer Engineering, Boulder CO 80309, USA\\
\authormark{4} NIST, Center for Nanoscale Science and Technology, 100 Bureau Dr, Gaithersburg, MD 20899, USA \\
\authormark{5} Maryland Nanocenter, University of Maryland, College Park, MD 20742, USA \\
\authormark{6} NIST, Time and Frequency Division, 325 Broadway, Boulder CO 80305, USA\\
\authormark{$\dagger$} Now with EPFL, Route Cantonale, 1015 Lausanne, Switzerland\\}

\email{\authormark{*}jeffrey.chiles@nist.gov} %% email address is required

% \homepage{http:...} %% author's URL, if desired

%%%%%%%%%%%%%%%%%%% abstract %%%%%%%%%%%%%%%%
%% [use \begin{abstract*}...\end{abstract*} if exempt from copyright]

\begin{abstract*}
The microscale integration of mid- and longwave-infrared photonics could enable the development of fieldable, robust chemical sensors, as well as highly efficient infrared frequency converters.  However, such technology would be defined by the choice of material platform, which immediately determines the strength and types of optical nonlinearities available, the optical transparency window, modal confinement, and physical robustness.  In this work, we demonstrate a new platform, suspended AlGaAs waveguides integrated on silicon, providing excellent performance in all of these metrics.  We demonstrate low propagation losses within a span of nearly two octaves (1.26 to 4.6 {\textmu}m) with exemplary performance of 0.45~dB/cm at $\lambda = 2.4$ {\textmu}m. We exploit the high nonlinearity of this platform to demonstrate 1560 nm-pumped second-harmonic generation and octave-spanning supercontinuum reaching out to 2.3~{\textmu}m with 3.4~pJ pump pulse energy.  With mid-IR pumping, we generate supercontinuum spanning from 2.3 to 6.5 {\textmu}m.  Finally, we demonstrate the versatility of the platform with mid-infrared passive devices such as low-loss 10~{\textmu}m-radius bends, compact power splitters with 96 $\pm$ 1\% efficiency and edge couplers with 3.0 $\pm$ 0.1 dB loss.  This platform has strong potential for multi-functional integrated photonic systems in the mid-IR.
\end{abstract*}

\counterwithout{figure}{section}
\counterwithout{figure}{subsection}

%%%%%%%%%%%%%%%%%%%%%%%%%%  body  %%%%%%%%%%%%%%%%%%%%%%%%%%
\section{Introduction}
\label{sec:introduction}
The mid- and longwave-infrared (mid-IR and LWIR) spectral regions from $\lambda = $ 3--8 and 8--15~{\textmu}m are critical spectral regions for sensitive spectroscopic analysis of a variety of physical compounds such as complex molecular solids, gaseous species and liquid mixtures \cite{griffiths2007fourier}. Optical frequency combs in particular have been used to achieve broadband spectroscopy with exquisite frequency resolution \cite{Adler2010,Schliesser2012,Coddington2016,Cossel2017a,Muraviev2018,Timmers2018,Picque2019}. However, broadband mid-IR comb technology is still maturing and nearly all systems would greatly benefit from increased spectral coverage, lower power operation and improved robustness. Consequently, there has been a consistent push over the past decade to transition mid-IR comb systems or sub-systems to compact and robust chip-scale platforms \cite{Cheng2012,Hugi2012,Yu2013,Shankar2013a,Kuyken2015,Penades2016a,Griffith2016,vasiliev2016chip,Kozak2018,Nader2018a,Ramirez2018,Stanton2019,Grassani2019,Sterczewski2019}. In particular, the small modal area and long propagation lengths of integrated nanophotonic waveguides motivate the development of integrated frequency converters to extend the spectral reach of combs often with negligible power burdens.

The successful development of integrated mid-IR photonic systems for these applications will ultimately depend on many factors, such as the material platform.  Only a small subset of materials have suitable optical transparency \cite{Soref2010}, and strong optical nonlinearities are also required for the generation or broadening of frequency combs in the mid-IR \cite{Griffith2016}.  While significant Kerr nonlinearity is present in silicon, germanium and chalcogenide materials, they lack intrinsic second-order optical nonlinearities for highly efficient frequency conversion \cite{Eyres2001,Muraviev2018,Timmers2018,Kowligy2018} and electro-optic modulation \cite{Chiles2014}. 

Alternatively, group III-V materials possess many desirable properties for multi-functional integrated photonic systems including a high refractive index, strong second- and third-order optical nonlinearities, and wide optical transparency windows into the LWIR.  A practical advantage of these materials is the ability to grow a chemically selective etch stop underneath a high-quality epitaxial device (donor) film, enabling wafer or chip-bonding film transfer techniques for heterogeneous integration \cite{Yablonovitch1990,Park2005}.  This has enabled high-index-contrast III-V waveguides on other substrates such as oxidized silicon and sapphire \cite{Dave2015,Hu2018,Honl2018,Chang2018,Zheng2018}.  However, to take full advantage of the broad transparency window supported by III-V semiconductors, it is necessary to pursue alternative geometries such as air-clad suspended waveguides.  But even this approach requires a degree of caution, as most materials readily form surface oxide layers that also introduce absorption. Undercut etching has been used to suspend GaAs waveguides engineered for mid-IR difference frequency generation \cite{Stievater2014}.  While this represents a promising step in the development of nonlinear mid-IR photonics with III-V materials, many issues remain, such as the propagation loss in the mid-IR region, atmospheric stability, coupling losses, and amenability to wafer-scale production.  Without thick cladding layers protecting the surface, suspended III-V waveguides become susceptible to several strong mid-IR and LWIR absorption bands \cite{sheibley1966infrared} from surface oxidation.

\begin{figure}[t]
\label{fig:overview}
\centering\includegraphics[width=13cm]{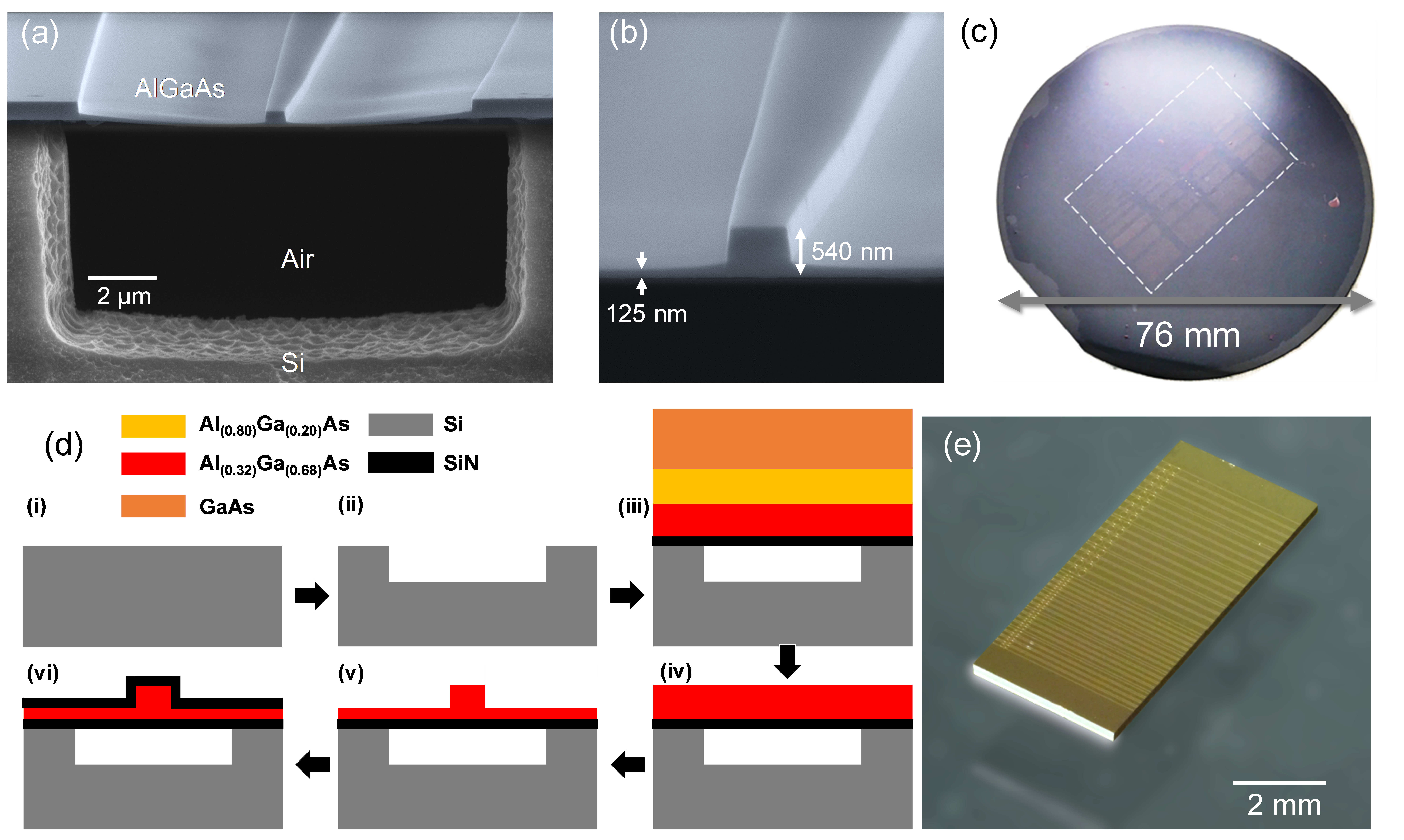}
\caption{The suspended AlGaAs on silicon platform. (a,b) Scanning electron microscope (SEM) images of a typical waveguide facet produced by dry etching. (c) Photograph of a processed wafer prior to die release (an AlGaAs film was bonded over the entire surface, but devices were only fabricated in the middle dashed portion to limit e-beam lithography write time.)  (d) Fabrication flow broken into six essential processing steps.  (e) Focus-stacked image of a microring resonator die after release from the wafer.}
\end{figure}

In this work, we present a new approach to air-clad mid-IR waveguides: suspended AlGaAs on silicon, which is shown schematically in Fig. \ref{fig:overview}. We employ direct-bonded membranes \cite{Chiles2013,Nader2018}, which provide superior mechanical stability and design flexibility thanks to the free choice of geometry and absence of stress-inducing perforations.  The issue of surface-oxide losses is addressed by passivating the AlGaAs surfaces with ultrathin films of sputtered silicon nitride (SiN), bringing the loss of waveguides from more than 100 dB/cm (without passivation) to <2.5 dB/cm (after passivation) through most of the mid-IR. The fabrication technique described in this work can also be applied to other III-V materials such as GaAs, GaP and GaN.  In the case of GaP and GaN, the larger bandgaps could enable even wider transparency windows.  However, due to the relative maturity of its device processing and material growth, we chose AlGaAs as the suspended material here.

The objective of this study is to provide a comprehensive demonstration of this platform's suitability for further development.  The work is divided as follows.  In Sec. \ref{sec:fabrication}, we describe the main fabrication approach to realize Al\textsubscript{0.32}Ga\textsubscript{0.68}As (bandgap of $\lambda = $ 681 nm) photonic waveguides on a conventional silicon 76 mm-diameter wafer (Fig. \ref{fig:overview}).  In Sec. \ref{sec:passives}, we report the design and characterization of several passive devices, including microring resonators, compact waveguide bends, inverted taper edge couplers and multimode interferometer (MMI) 1 $\times$ 2 splitter junctions. In Sec. \ref{sec:nonlinear}, we utilize dispersion-engineered waveguides to achieve second-harmonic generation and supercontinuum generation from femtosecond laser sources at low pulse energies, and we compare the results for near-IR and mid-IR pumping.  We conclude with a discussion of the platform's merits and prospects for future research.  Supplement 1 discusses mid-IR losses from surface states and passivation layers, damage threshold measurements and physical robustness, and the measurement setups.

%overview fig

\section{Results}

\subsection{Fabrication}
\label{sec:fabrication}
A schematic representation of the fabrication process is provided in Fig. \ref{fig:overview}(d).  Throughout this project, several fabrication runs were conducted with variations on the main fabrication flow.  The illustrated process captures the essential details of the best-performing variation, on which most of the results of this work are focused.  The fabrication process consists of (i) cleaning a silicon handle wafer, (ii) etching trench features where waveguides will be situated, (iii) bonding an epitaxial III-V wafer to the handle (a SiN passivation layer has been applied prior to the bond), (iv) chemically removing the III-V wafer substrate (GaAs) and etch-stop layer (Al\textsubscript{0.8}Ga\textsubscript{0.2}As), (v) electron-beam lithography and plasma etching of the ridge waveguide features over the trenched areas, and (vi) final oxide strip and top surface passivation with SiN.  Afterward, individual dies were released from the wafer simultaneously using deep reactive ion etching (DRIE) through the silicon handle wafer.  The dies were then annealed at $300^{\circ}$C in a nitrogen environment, which reduced losses from N-H bond absorption; further discussion is available in Supplement 1.  The devices have a nominal core thickness of 540 nm and a slab thickness of 125 nm (Fig. \ref{fig:overview}(b)).  The trenches underneath the waveguides are 6 {\textmu}m deep and about 11 {\textmu}m wide.  The SiN barriers on the bottom and top surfaces of the membrane are 10 and 20 nm thick, respectively.  No die-level processing was necessary, enabling parallel fabrication of dozens of chips without any polishing and cleaving (which is generally required for sapphire or group III-V substrates).  This is a key strength of utilizing silicon substrates.  

%ring fig shows the microring resonator results and analysis
\begin{figure}[t]
\centering\includegraphics[width=13cm]{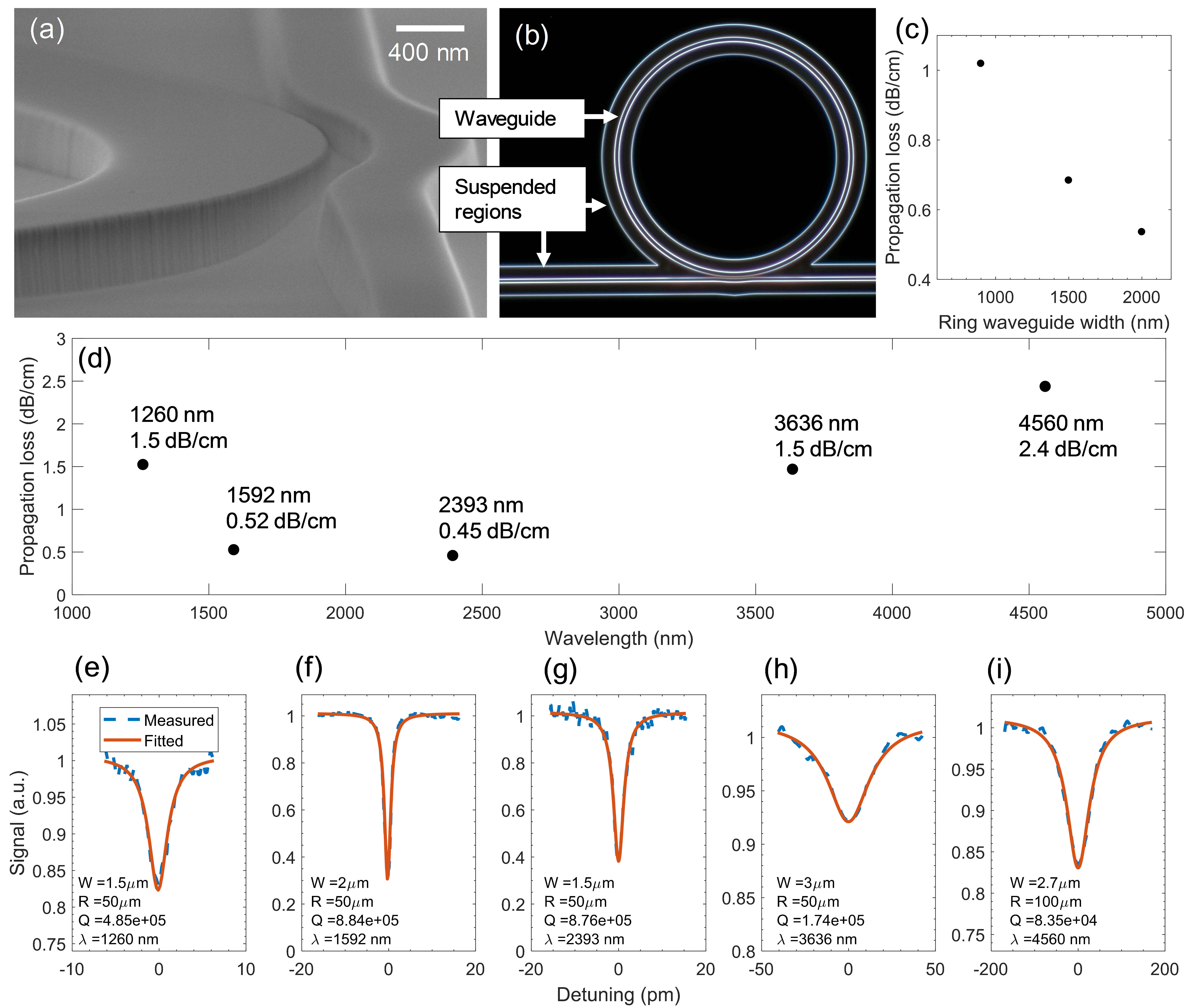}
\caption{Characterization of microring resonators.  (a) SEM image of a resonator, showing pulley-coupler region and sidewall roughness of a ring.  (b) Darkfield optical micrograph of the top view of a fabricated resonator.  Suspended regions encompass all waveguide features on the substrate. (c) Ring waveguide width vs. propagation loss for resonances taken at $\lambda = $ 1564, 1556 and 1592 nm, from left to right.  (d) Compiled propagation loss data vs wavelength. The first two data points (1260 and 1592 nm) use the intrinsic rather than loaded \textit{Q}.  For the others, the signal was AC-coupled (giving uncertain extinction ratio), so we used loaded \textit{Q}. (e-i) Experimental and fitted traces for the resonance considered in each datapoint of subplot (d).  W: width of ring waveguide; R: radius of ring resonator; Q: loaded quality factor.}
\label{fig:ringfig}
\end{figure}

\subsection{Passive Measurements}
\label{sec:passives}

High-performance passive devices are required to eventually integrate multiple components on one chip, such as nonlinear frequency converters, modulators and detectors.  The ability to simultaneously realize precisely dispersion-engineered waveguides with tight bending radii, small mode volumes and low coupling losses is not guaranteed in all waveguide platforms.  To this end, we designed, fabricated and tested several key passive elements including microring resonators, bends, 1 $\times$ 2 MMI power splitters, and input/output (edge) couplers.  First, we consider the characterization of suspended AlGaAs microring resonators over a broad wavelength range from the near- to mid-IR.  We show that the passivation approach was effective at suppressing the inherently high absorption loss of bare AlGaAs surfaces, achieving losses below 2.5 dB/cm out to $\lambda =$ 4.6 {\textmu}m.  An analysis focused on specific loss contributions in the mid-IR, including surface states of unpassivated waveguides, can be found in Supplement 1.

\subsubsection{Microring resonators and propagation loss}
\label{sec:microrings}

Microring resonators provide a convenient gauge for the waveguiding performance of any integrated photonic platform.  Since we are interested in nonlinear applications involving pump wavelengths across the near-IR and mid-IR, we examined the performance of several resonator devices at $\lambda =$ 1.26, 1.59, 2.39, 3.64 and 4.56 {\textmu}m.  For all wavelengths, light was aligned to the quasi-transverse-electric (TE) mode of the waveguides. Further discussion of the measurement setups and laser systems is available in Supplement 1. 

The results are collected in Fig. \ref{fig:ringfig}.  The broadband spectral dependence of propagation loss is shown in Fig. \ref{fig:ringfig}(d).  The relevant device parameters (ring waveguide width, ring radius, etc.) and measurement results are shown for each resonance in Fig. \ref{fig:ringfig}(e-i).  The resonances in this figure each represent the highest quality factor (\textit{Q}) observed for that wavelength point. The maximum loaded quality factor of $8.8 \times 10^5$ is achieved in the 2 {\textmu}m-wide ring device at $\lambda =$ 1592 nm.  Factoring out the coupling loss of the ring (conservatively assuming an under-coupled condition), this corresponds to an intrinsic \textit{Q} of $1.1 \times 10^6$, giving a waveguide propagation loss of 0.52 dB/cm.  Going to longer wavelengths, we measured propagation losses of 0.45, 1.5 and 2.4 dB/cm for $\lambda =$ 2.39, 3.64 and 4.56 {\textmu}m, respectively. It can be seen that the propagation loss exhibits a trough between $\lambda =$ 1500 - 2500 nm, with a sharply rising loss on the short-wavelength side and a slow increase toward longer wavelengths. The blue-side increase can readily be attributed to the well-known phenomenon of surface-state absorption observed in GaAs/AlGaAs waveguides, which increases sharply at wavelengths near $\lambda = 1$ ~{\textmu}m \cite{kaminska1987el2}. The fabrication process was not optimized to enhance the performance in this region, but specific treatments are possible if low-loss operation is required here \cite{Chang1995}.  The loss observed from $\lambda =$ 1500 - 2500 nm is mostly scattering loss. The trend of ring waveguide width vs. loss is plotted in Fig. \ref{fig:ringfig}(c). The significant negative slope with respect to the waveguide width, and the fact that the loss has not yet saturated at wider widths, indicate there is room for improvement in the sidewall roughness of the waveguides.  Figure \ref{fig:ringfig}(a) shows evidence of sidewall corrugations resulting from roughness in the lithography that is transferred directly to the device, indicated by striations that are uniform along the vertical axis, but varying in position in the width axis of the waveguide.  Multi-pass electron-beam lithography \cite{Ji2017} or resist reflow can be employed to mitigate this effect.  We discuss long-wavelength losses and possible origins in Supplement 1, including a narrow peak in absorption near $\lambda = 3$ {\textmu}m which could not be analyzed with ring resonators.  As a final note to this section, the ring resonance data was taken more than two months after fabrication of the chips, proving the effectiveness of the passivation layer at preventing losses from oxidation over time.  

%ring scans
\begin{figure}[t]

\centering\includegraphics[width=13cm]{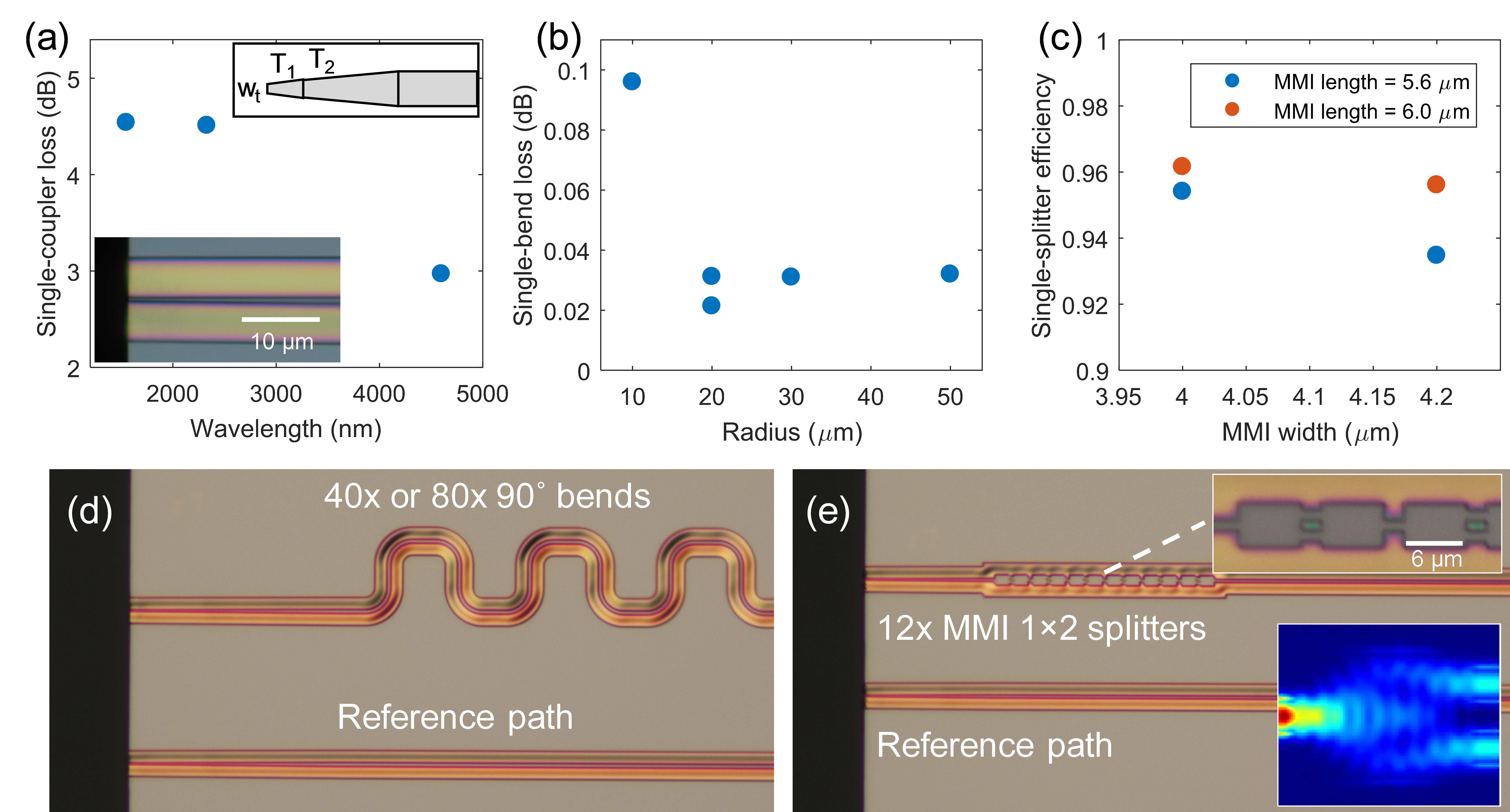}
\caption{Experimental characterization of passive suspended AlGaAs components. (a) Loss-per-edge-coupler at different wavelengths. Top inset: inverted taper geometry, showing tip width \textit{w\textsubscript{t}}, and taper sections \textit{T\textsubscript{1}} and \textit{T\textsubscript{2}} (see Table \ref{tab:coupler}). Bottom inset: optical micrograph of an inverted taper edge coupler. (b) Bend radius vs. single-bend loss at $\lambda = 4.6$ {\textmu}m for a 1.4 {\textmu}m-wide waveguide. (c) MMI power splitter efficiency for several variations on the length and width of the multimode propagation section. (d,e) Optical micrographs of cutback structures used to measure single-bend and single-splitter loss, respectively.  Top inset of (e): zoom view of consecutive MMIs.  Bottom inset of (e): top view of the simulated optical intensity of the designed MMI splitter.}
\label{fig:pass}
\end{figure}

\subsubsection{Coupling loss}

In order to achieve efficient frequency conversion with integrated photonic devices, it is critical to have low-loss and broadband on- and off-chip couplers.  We pursued inverted taper couplers to meet these goals. The facets are prepared in parallel via plasma etching. An optical micrograph of a typical edge coupler is shown in the lower inset of Fig. \ref{fig:pass}(a). They employ two stages of tapering (upper inset of Fig. \ref{fig:pass}(a)), which reduces the total length . We consider several couplers with corresponding wavelengths of interest to this project: 1.55, 2.3 and 4.6 {\textmu}m. The results are shown in Fig. \ref{fig:pass}(a), and the coupler geometrical parameters are given in Table \ref{tab:coupler}. In many cases, the taper length did not factor significantly into the performance, but it is included for completeness.  The lowest coupling loss of 3.0 $\pm$ 0.1 dB is achieved at $\lambda = 4.6$ {\textmu}m, where this value corrects for the expected propagation loss of 0.2 dB in a 1 mm long chip at this wavelength.  Losses are reported with respect to the reference transmission of light through both aspheric lenses with no chip.  The increase in loss toward shorter wavelengths is expected due to the slab portion constraining the modal extent in the vertical direction.  Fully-suspended taper geometries may be employed in the future to enhance the performance at shorter wavelengths \cite{Nader2018}.  

\newcommand{\ra}[1]{\renewcommand{\arraystretch}{#1}} % more space between rows
\begin{table}[t]\centering
\caption{Edge coupler geometries.}
\small
\ra{1.2}
\begin{tabular}{@{}cccc@{}}
\toprule
$\lambda$ (nm) & Tip width (\textit{w\textsubscript{t}})  & Taper 1 (\textit{T\textsubscript{1}}) & Taper 2 (\textit{T\textsubscript{2}})\\
\midrule
1550 & 230 nm & to 500 nm over 36 {\textmu}m & to 1000 nm over 36 {\textmu}m  \\
2320 & 210 nm & to 650 nm over 36 {\textmu}m &to 1000 nm over 31 {\textmu}m \\
4600 & 300 nm & to 550 nm over 15 {\textmu}m & to 1200 nm over 124 {\textmu}m \\

\bottomrule
\end{tabular}
    \label{tab:coupler}
\end{table}

\subsubsection{Bending loss}

Waveguide bends are an essential element to almost any integrated photonic system, but some material platforms intrinsically limit bending performance based on the lateral or horizontal index contrast, or the typical mode volume required to realize low-loss waveguides.  For waveguides which take advantage of strong modal confinement to avoid cladding losses \cite{Miller2017}, relatively large, graded-curvature bends are required to avoid modal mixing.  In other platforms where limited index contrast is available \cite{Ramirez2018}, bends are also constrained by leakage losses. With suspended AlGaAs on silicon, however, a strong index contrast can be achieved on all sides of the waveguide, enabling small mode volumes and compact bends.  We utilized the structures shown in Fig. \ref{fig:pass}(d) to test the bending performance.  They consist of a reference path with only a straight waveguide, and an adjacent path containing forty or eighty 90$^\circ$ bends, each separated by 10 {\textmu}m of straight waveguide length, to properly assess the mode transition loss (each bend having two straight-to-bend transitions).  Propagation losses are also part of the total loss, though the effect of the length difference is generally not significant compared to the transition loss.  These structures were repeated over an array of several widths and radii and tested at both $\lambda = 3.4$ and 4.6 {\textmu}m.  The bend loss dependence on radius at $\lambda = 4.6 $ {\textmu}m is shown in Fig. \ref{fig:pass}(b), with a typical value of 0.03 dB loss per bend above radii of 20 {\textmu}m.  The waveguide width was 1.4 {\textmu}m.  At $\lambda = 3.4$ {\textmu}m, 0.03 dB loss per bend is observed for a radius of 20 {\textmu}m (1.4 {\textmu}m waveguide width), and 0.06 dB loss per bend at 10 {\textmu}m radius (1.0 {\textmu}m waveguide width).  For reference, the simulated loss of one 10 {\textmu}m radius bend is 0.04 dB, in good agreement considering the extra loss is likely from sidewall scattering.

\subsubsection{Multimode interferometer splitters}

We also investigated $1\times2$ power splitters, another important building block for multi-functional photonic systems in the mid-IR and LWIR regions. We used MMI-type splitters for compactness, as opposed to adiabatic or directional coupler-based designs.  The simulated design used the following parameters: MMI length of 6 {\textmu}m, MMI width of 4 {\textmu}m, port waveguide width of 1 {\textmu}m, and port waveguide center-to-center gap of 2 {\textmu}m.  The simulated efficiency was 92 \%.  The simulated optical intensity from a top view is shown in the lower inset of Fig. \ref{fig:pass}(e).  We characterized the devices with the structures in Fig. \ref{fig:pass}(e).  They consist of a reference path with only a straight waveguide, and an adjacent path containing 12 consecutive $1\times2$ or $2\times1$ devices. Comparing their transmission at $\lambda = 3.4$ {\textmu}m, we observed an experimental device efficiency of 96 $\pm$ 1\%.  We also tested several variations in design dimensions, and the overall performance in Fig.~\ref{fig:pass}(c) was robust to variations of several hundred nm in the MMI length and width parameters. The difference in performance between the simulated and experimental results could be partly a result of unexpected deviations from the design (such as rounded features from lithography and etching) and partly from error in the reference path normalization.

%include table at bottom

\subsection{Nonlinear measurements}
\label{sec:nonlinear}
Next, we focus on the nonlinear optical characterization.  This includes two cases: near-IR-pumped and mid-IR-pumped supercontinuum generation. A detailed discussion of the measurement setups in both cases can be found in Supplement 1, as well as a discussion of the damage thresholds observed during experiments.

%telecom scg
\begin{figure}[t]

\centering\includegraphics[width=13cm]{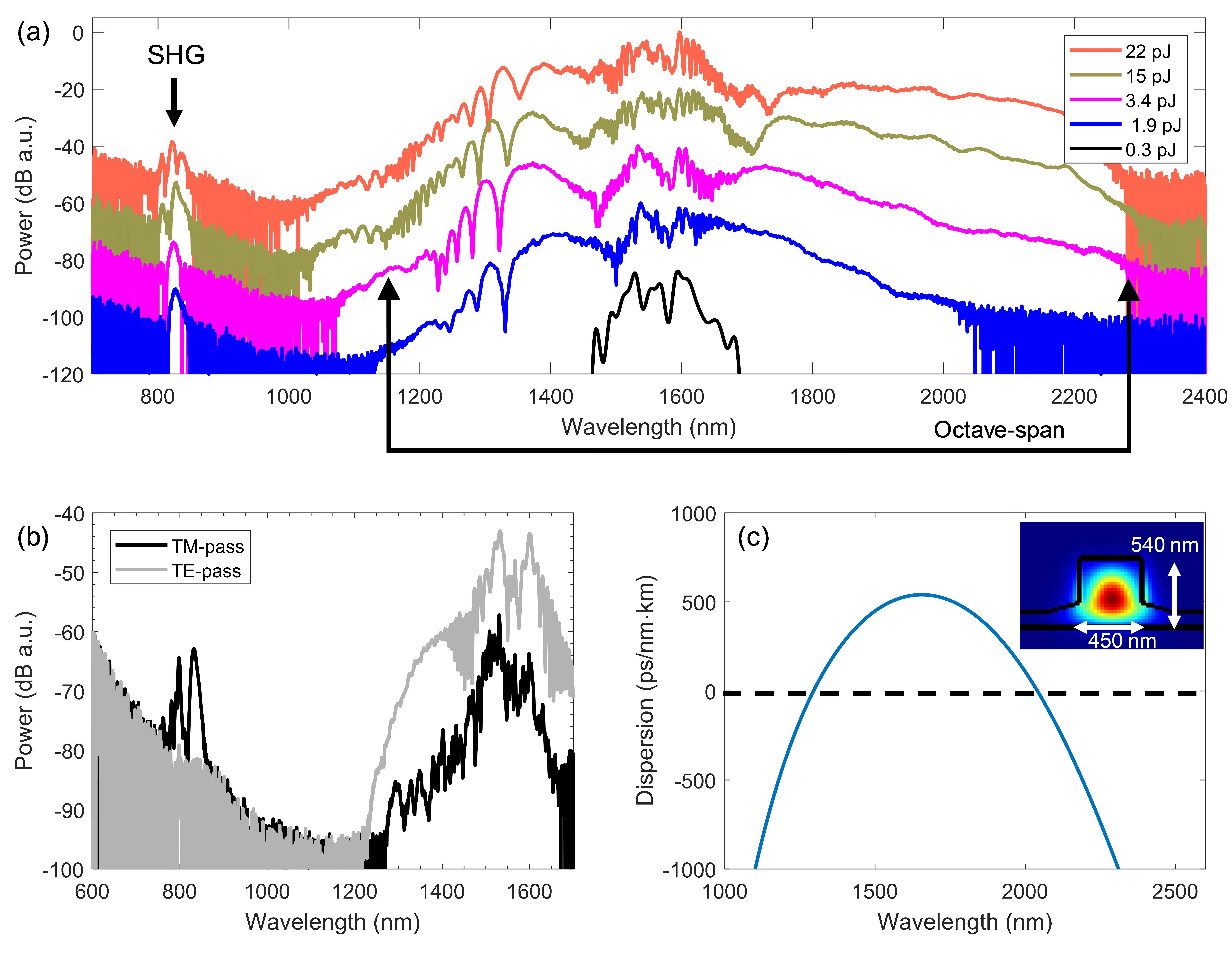}
\caption{Supercontinuum generation from a 1560 nm pump. (a) Experimentally measured spectra for different waveguide-coupled pulse energies.  Octave-spanning bandwidth is highlighted for the case of 3.4 pJ. Trace-to-trace offset is 20 dB. (b) Polarized supercontinuum output of a similar waveguide for the TE- and TM-pass cases, showing suppression of the second-harmonic peak for TE-pass. (c) Simulated waveguide GVD. Inset: intensity profile for the mode under consideration.}
\label{fig:telecom_scg}
\end{figure}

\subsubsection{Near-IR-pumped supercontinuum generation}
\label{sec:telecom_scg}

First, we consider suspended AlGaAs waveguides that are dispersion-engineered to achieve anomalous group velocity dispersion (GVD) near the pump wavelength of 1560 nm (Fig. \ref{fig:telecom_scg}(c)). The waveguides have a core width of 450 nm, an effective mode area of 0.22 {\textmu}m$^2$, and a length of 4 mm.  Due to a slight curvature of the etch profile near the waveguide core (Fig. \ref{fig:overview}(b)), the mode simulations incorporate a sloped region on the sidewalls as shown in the inset of Fig. \ref{fig:telecom_scg}(c).  The pump is an amplified and compressed erbium-fiber oscillator with a repetition rate of 160 MHz, a pulse width of 61 fs, and a center wavelength of 1560 nm. Light was coupled in to the chip through an aspheric lens, and out through a single-mode lensed fiber. Figure \ref{fig:telecom_scg}(a) shows the output spectral evolution as the waveguide-coupled pump pulse energy is varied.  At 300~fJ pump pulse energy, the waveguide output spectrum exhibited no noticeable broadening compared to the input comb spectrum.  The onset of supercontinuum generation was observed near 1.9 pJ, and the spectrum was noticeably saturated by three-photon absorption (3PA) at 15 pJ.  We observed octave-spanning supercontinuum generation (at $-$45 dB level with respect to pump intensity) at a pulse energy of 3.4 pJ (average waveguide-coupled power of 0.5 mW). We also observed a second-harmonic generation (SHG) peak near 830 nm for almost all input power levels. This results from phase-mismatched conversion over short lengths on the chip and has been observed in other waveguide platforms possessing nonzero $\chi^{(2)}$ \cite{Hickstein2017,Yu2019}.  We confirmed this by polarizing the output spectrum in vertical (TM) or horizontal (TE) directions in Fig. \ref{fig:telecom_scg}(b). The peak was TM-polarized.  This polarization relationship (TE pump, TM harmonic) is expected for GaAs or AlGaAs waveguides propagating along the [0$\bar{1}\bar{1}$] axis. The low pulse energies required to initiate soliton fission in these waveguides and the simultaneous production of single-spatial-mode supercontinuum and second-harmonic generation could benefit numerous applications, including frequency combs that could be self-referenced with no amplifiers following the oscillator, or very high repetition-rate systems \cite{Klenner2016a,Shoji2016,Carlson2018}. Significantly stronger SHG can readily be achieved in the future through the use of quasi-phase-matching \cite{Rao2017,Roux2017}.  The short-wavelength supercontinuum could also be enhanced via treatments to reduce the short-wavelength losses from surface states on the AlGaAs membrane \cite{Chang1995}, which we assume to be the current limitation to the bandwidth.

%mid ir scg fig
\begin{figure}[t]

\centering\includegraphics[width=13cm]{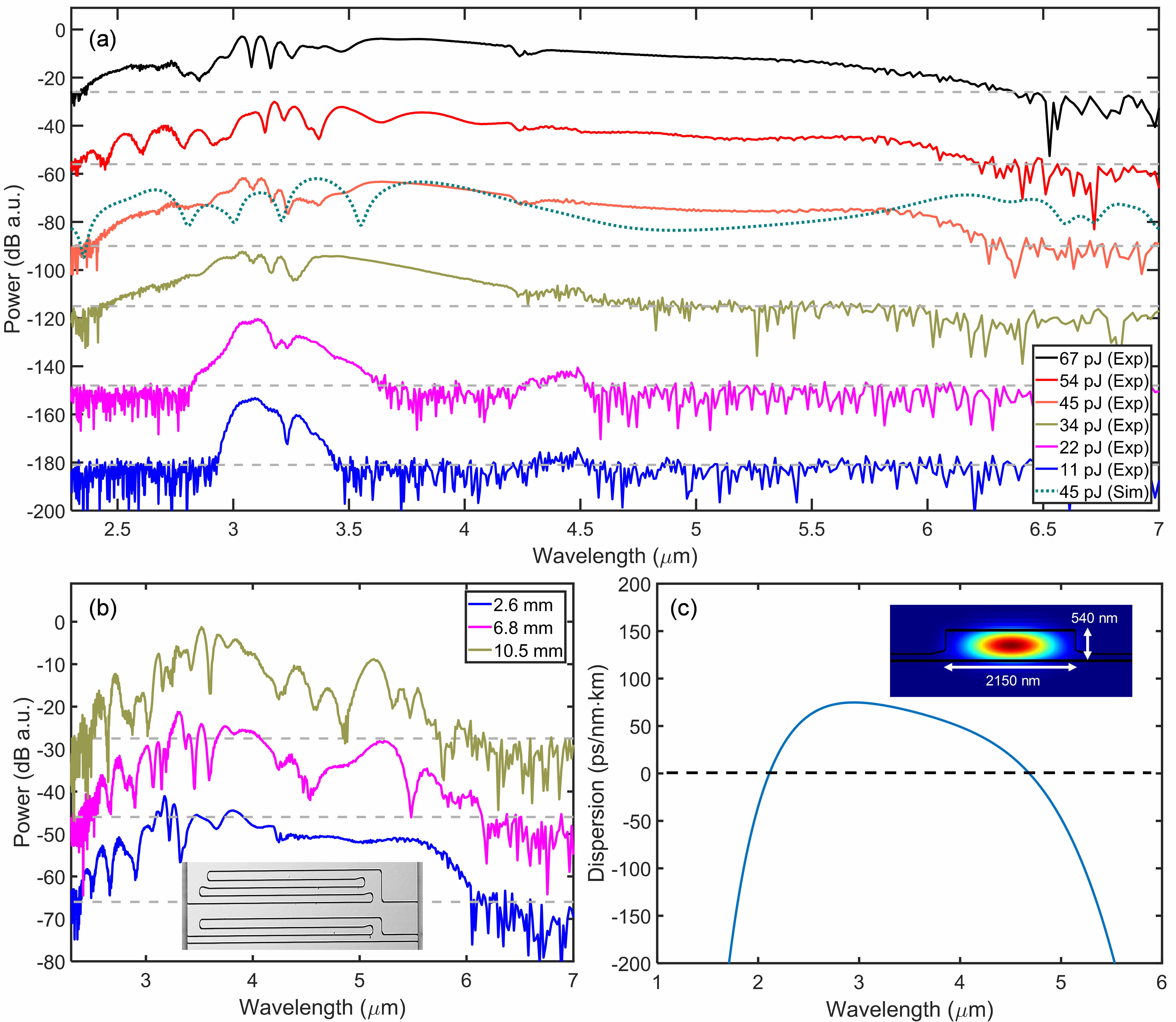}
\caption{(a) Supercontinuum generation from a 3060 nm pump;  experimentally measured spectra for different waveguide-coupled pulse energies (solid lines), and simulated spectrum at 45 pJ waveguide-coupled pulse energy (dotted line). Trace-to-trace offset: 30 dB. (b) Supercontinuum output for various total waveguide lengths at 67 pJ pulse energy. Trace-to-trace offset: 20 dB.  Inset: optical micrograph of a paperclip structure used for length variations.  Dashed grey lines in (a,b) indicate approximate noise floor for each trace. (c) Simulated waveguide GVD. Inset: intensity profile for the mode under consideration.}
\label{fig:midir_scg}
\end{figure}

\subsubsection{Mid-IR-pumped supercontinuum generation}

We now consider the broadening of pulsed lasers in the mid-IR region.  Here, the pump was produced via difference frequency generation of an amplified and broadened erbium-fiber comb output. The nominal pulse width was 80 fs, the repetition rate was 100 MHz, and the center wavelength was 3.06 {\textmu}m \cite{Cruz2015}.  We focused on a waveguide geometry with a core width of 2.15 {\textmu}m, an effective mode area of 1.2 {\textmu}m$^2$, and a nominal length of 2.3 mm, excluding tapered regions. The GVD and intensity mode profile are depicted in Fig. \ref{fig:midir_scg}(c), showing a much flatter and broader region of anomalous dispersion compared to the 1550~nm case (owing to the strongly reduced bulk GVD contribution in the mid-IR for AlGaAs).  The waveguides were measured by coupling the pump light in the TE-mode orientation and analyzing the output spectrum in a Fourier-transform infrared spectrometer (FTIR).  The light was coupled out through an aspheric lens aligned to collimate the long-wave side of the spectrum. More details and a schematic describing the setup can be found in Supplement~1.  Figure \ref{fig:midir_scg}(a) shows the measured spectra at different waveguide-coupled pump pulse energies. At the lowest energy of 11 pJ, the pump exhibited no significant broadening. Supercontinuum generation occurred at 45 pJ, showing a smooth, mostly unstructured spectrum spanning out to 6200 nm with a small dispersive wave peak near 6000 nm.  At higher pump powers, the spectrum flattened out and broadened slightly.  We also investigated variations on the waveguide length in Fig. \ref{fig:midir_scg}(b). A fixed waveguide-coupled pulse energy of 67 pJ was used here. As expected, longer waveguides resulted in the generation of more solitons, giving a more structured spectrum as the pulse progressed to a final length of 10.5 mm.

Returning to Fig. \ref{fig:midir_scg}(a), we compare the experimental vs. simulated results for 45 pJ of waveguide-coupled pulse energy.  The simulations use an  \textit{n}\textsuperscript{(2)} of 0.9$\times 10^{-17}$ m$^{2}$/W and a uniform propagation loss of 10 dB/cm. Note that this loss is actually confined to a narrow peak near $\lambda = 3$ {\textmu}m from N-H bond absorption; more details and analysis on it can be found in Supplement 1.  We used the harmonic oscillator model from Ref. \cite{Terry1991} to calculate the AlGaAs material GVD. The structure of the simulated spectrum largely agrees with the experiment. The simulation produces a dispersive wave near $\lambda = 6$ {\textmu}m with a smooth region connecting it to the pump. However, broadening beyond $\lambda = 6.5$ {\textmu}m was not experimentally observed in this geometry, and a simulated dispersive wave expected at $\lambda = 1.3$ {\textmu}m was also not observed.  The absence of the short-wave dispersive wave may be due to the reduced sensitivity of the FTIR near the limit of its detection range, but the difference in long-wave structure is more likely from a discrepancy in the material dispersion, since extrapolation was required for wavelengths longer than 830 nm.  The location of the dispersive wave is strongly dependent on the long-wave zero-crossing in the GVD curve, so even a small difference could result in the observed dispersive wave shift. This could be resolved with direct measurements of the refractive index of this AlGaAs composition in the mid-IR using spectroscopic ellipsometry.

In principle, much larger broadening in the long-wave spectrum is expected for waveguides with modified dispersion profiles, such as deeper etches or thicker core regions.  The thickness in this work was chosen as a compromise enabling broadband anomalous dispersion to be engineered from $\lambda =$ 1.5 to 4.7 {\textmu}m depending on the waveguide width.  As a final note, in contrast to the 1560 nm pumping case, no second-harmonic generation was observed for 3060 nm pumping, most likely due to the larger phase-mismatch between the pump and signal.

\section{Conclusion}

In this work, we presented a new integrated photonic platform, suspended AlGaAs on silicon.  By using directly bonded AlGaAs membranes on pre-etched trenches in a silicon substrate, we have overcome many obstacles to the adoption of III-V materials for integrated photonics in the mid-IR.  With this approach, multi-functional devices leveraging $\chi^{(2)}$ and $\chi^{(3)}$ nonlinearities can be fabricated reliably at the wafer-scale with high coupling efficiency and low propagation losses.  We show the all-around strengths of this platform through a series of linear and nonlinear experiments.  The relatively wide bandgap of this AlGaAs composition has enabled us to use near-IR pumps to achieve second-harmonic generation and octave-spanning supercontinuum generation in dispersion-engineered AlGaAs waveguides at remarkably low waveguide-coupled pulse energies of 3.4~pJ. We also realized octave-spanning supercontinuum generation in the mid-IR, from $\lambda =2.3 - 6.5$ {\textmu}m. High-quality microring resonators were fabricated, and a loss of 0.45~dB/cm was realized at a wavelength of 2.4~{\textmu}m.  A minimum coupling loss of 3.0 $\pm$ 0.1 dB/facet was observed at $\lambda = 4.6$ {\textmu}m.  Low-loss passive elements including compact waveguide bends (0.06 dB loss per 90$^{\circ}$ bend with 10 {\textmu}m radius) and $1 \times 2$ MMI power splitter junctions with only 6 {\textmu}m total length and 96 $\pm$ 1\%  power efficiency have also been demonstrated in the mid-IR. Crucial to the practical development of this platform, we characterized the performance of unpassivated AlGaAs waveguides and showed that proper passivation is essential to low-loss operation in the mid-IR.  We developed SiN passivation allowing waveguides to be operated in air for long time periods (most loss values were measured >2 months after fabrication).  All of this has been achieved at the wafer-scale with no required die-level processing, a valuable practical advantage over native III-V platforms. The combination of a wide optical transparency window, strong $\chi^{(2)}$ and $\chi^{(3)}$ nonlinearities, and the bandgap of 681 nm may enable ultra-efficient frequency converters bridging the gap from the near-IR to the mid-IR. Furthermore, the observed ultra-low thresholds for optical nonlinearities, combined with the low propagation losses realized in this work may point to new opportunities for on-chip quantum photonics \cite{Cabello2012,Hamel2014} at 1550 nm and beyond.
Suspended AlGaAs on silicon is a high-performance platform for multi-functional integrated photonics with both passive and nonlinear systems.

\section*{Funding}
Partial funding provided by: NIST, DARPA through the SCOUT program, and AFOSR under contract no. FA9550-15-1-0506.

\section*{Acknowledgment}
We thank Thorlabs, Inc. for the use of their MLQD4550 QCL product.  We also acknowledge FLIR Systems, Inc. for the use of their InSb camera during part of the experiments.  We thank the following individuals at NIST Boulder: Esther Baumann, Gabriel Colacion and Jacob Friedlein for assistance configuring the 1560 nm fiber comb system for nonlinear experiments, Kimberly Briggman for the FTIR measurements, and Paul Blanchard for assistance with SEM imaging.

This work is an official contribution of the National Institute of Standards and Technology, not subject to copyright in the United States.  Product disclaimer: Any mention of commercial products is for information only; it does not imply recommendation or endorsement by NIST.

See Supplement 1 for supporting content.

\section*{Disclosures}
The authors declare that there are no conflicts of interest related to this article.

%%%%%%%%%% If using BibTeX:
\bibliography{OSA-template}

\newpage

\section{Supplementary Materials}

This document provides supplementary information to ``Multi-functional integrated photonics in the mid-infrared with suspended AlGaAs on silicon.''  In Section \ref{sec:unpassivated}, we discuss experimental results for unpassivated waveguides.  In Section \ref{sec:mirloss}, we analyze sources of propagation loss in the mid-IR. In Section \ref{sec:damage}, we discuss the damage threshold and physical robustness of suspended AlGaAs waveguides.  In Section \ref{sec:setups} we elaborate on the measurement setups for the passive and nonlinear characterization.

\subsection{Characterization of unpassivated waveguides}
\label{sec:unpassivated}

If no passivation is used on the AlGaAs waveguide surfaces, it will very rapidly form a native oxide layer consisting of various phases of arsenic, gallium and aluminum oxides.  In the near-IR, the resulting losses of this reconstruction layer are noticeable on the level of 1 dB/cm for nanoscale waveguides \cite{Guha2017}.  However, the effects of this layer on photonic devices have not yet been studied in the mid-IR.  We tested several devices without SiN passivation to better understand its effects on these waveguides.  The results are summarized in Table S\ref{tab:unpassivated}.  In the first test, an unpassivated waveguide from an earlier fabrication run (core thickness 350 nm, waveguide width of 2.2 {\textmu}m) was tested via cutback at $\lambda = 3.4$ {\textmu}m to find the propagation loss once the surface was fully oxidized (after weeks in atmosphere).  The loss was so high that it could only be measured with 0.9 mm propagation length difference using a microbolometer array to resolve the faint out-coupled modes. A propagation loss > 100 dB/cm was observed.  Next, we examined a different waveguide (core thickness 540 nm, waveguide width of 2.2 {\textmu}m) which was only passivated on the bottom surface.  We stripped the top oxide using ammonium hydroxide and hydrochloric acid, then measured the loss roughly 1 hour later.  We recorded a loss of 16 dB/cm.  Finally, we examined the loss at $\lambda = 4.6$ {\textmu}m using ring resonator measurements.  The device under consideration had a core thickness of 540 nm, a ring waveguide width of 2.7 {\textmu}m, and only bottom-side passivation similar to the previous example.  A loss of 5.5 dB/cm was observed about 30 minutes after the top-oxide strip.  

\begin{table}[b]\centering
\caption{Summary of loss measurements for unpassivated waveguides.}
\small

\begin{tabular}{@{}ccccc@{}}
\toprule
$\lambda$ (nm) & Loss & Passivation & Time delay after oxide strip &  Method \\
\midrule
3400 & >100 dB/cm & none & weeks & cutback \\
3400 & 16 dB/cm & bottom side only (SiN) & 1 hr & cutback \\
4560 & 5.5 dB/cm & bottom side only (SiN) & 30 min & ring resonator \\

\bottomrule
\end{tabular}
    \label{tab:unpassivated}
\end{table}

A few conclusions may be drawn from these data.  The first is that unpassivated AlGaAs waveguides exhibit substantial absorption losses near $\lambda = 3.4$ {\textmu}m.  Secondly, it does not take much time for the losses to increase to unacceptable levels, meaning that even laboratory-scale demonstrations of similar platforms should count on implementing passivation prior to measuring devices.  Finally, the lower loss observed at $\lambda = 4.6$ {\textmu}m suggests that there is significant structure to the absorbance spectrum of the surface states, although it is still problematic well beyond the spike in loss observed near $\lambda = 3.4$ {\textmu}m.  This absorbance feature is worthy of future study to highlight its chemical origins, since it does not correspond to any previously observed feature in the absorption spectra of gallium or arsenic oxides \cite{sheibley1966infrared}.  Note that in Ref. \cite{sheibley1966infrared}, the absorption feature near $\lambda = 3$ {\textmu}m is due to water contamination in the samples, and should not be interpreted as an intrinsic feature of the materials.

\subsection{Mid-IR loss contributions}
\label{sec:mirloss}

In the main manuscript, we characterized the performance of microring resonators over a spectrum spanning from $\lambda = 1.26 - 4.6$ {\textmu}m. In this section of the Supplement, we consider potential contributions to loss in the mid-IR. Free-carrier absorption could be a factor, given that it generally exhibits loss that increases with wavelength.  However, reasonable estimates based on an intrinsic defect density of $1\times10^{15}cm^{-3}$ put the estimated free-carrier absorption at $\lambda = 5$ {\textmu}m at a value of $< 0.1$ dB/cm \cite{Spitzer1959}. Another potential source of loss is the SiN passivation layer, with a total thickness of 30 nm (10 nm on bottom surface, 20 nm on top).  First, we measured via prism coupling a bulk material loss of < 2 dB/cm at $\lambda = 1550$ nm for the SiN film by itself, indicating reasonably good film quality.  To test the performance in the mid-IR, we performed a differential measurement by depositing 20 nm of additional SiN material on a spiral waveguide cutback with known propagation loss.  Measuring at $\lambda = 3.4$ {\textmu}m, we observed an increase in propagation loss of roughly 1-2 dB/cm.  Since the optical mode overlap in the nitride film is only 2\%, the equivalent bulk SiN loss at $\lambda = 3.4$ {\textmu}m is on the order of 50-100 dB/cm.  

%3p1 loss fig
\begin{figure}[t]

\centering\includegraphics[width=13cm]{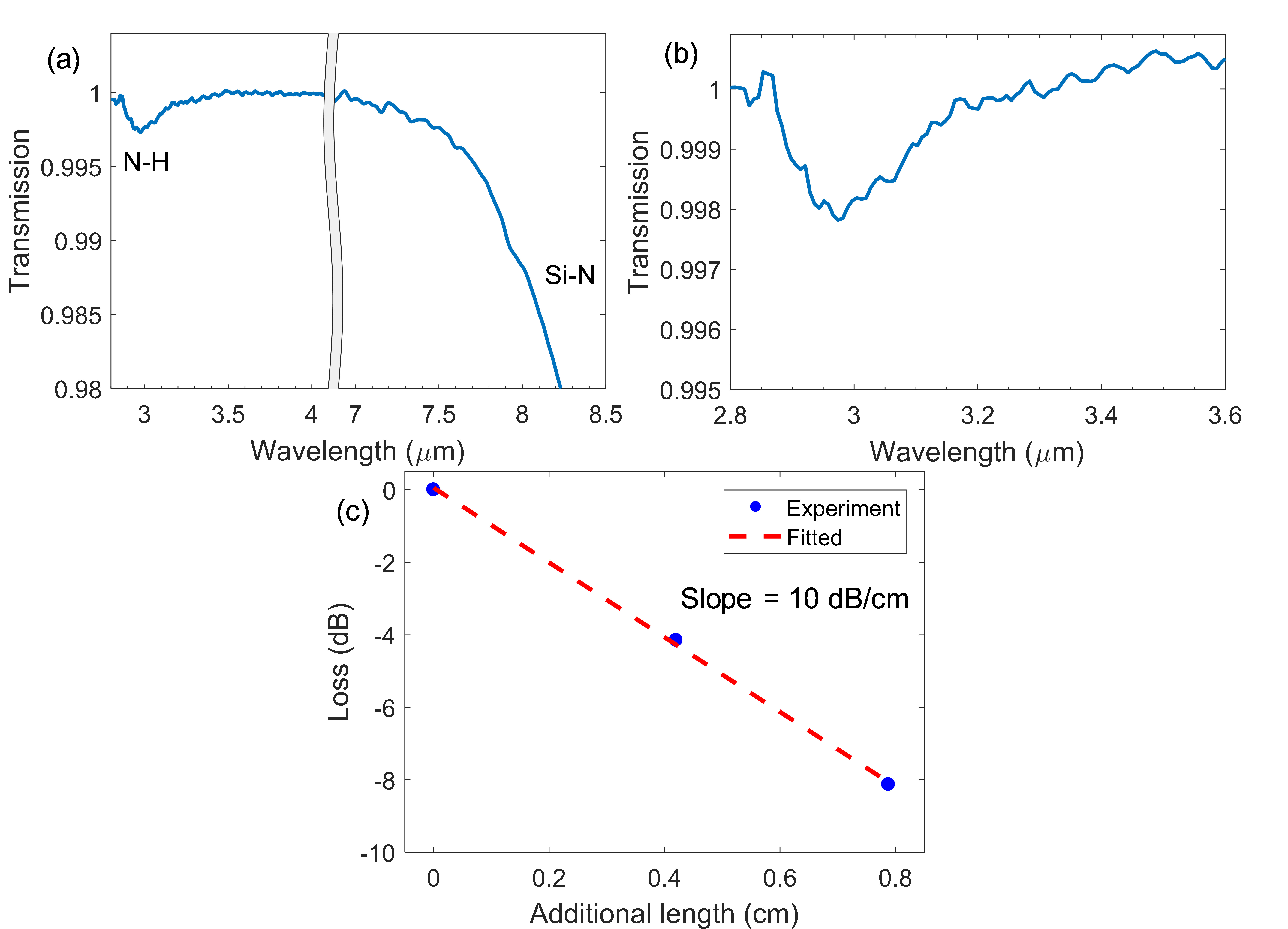}
\caption{(a) FTIR measurement of a thin sputtered SiN film, showing transmission dips from N-H and Si-N absorption features. (b) Zoom of the N-H absorption feature. (c) Measurement of the propagation loss at $\lambda = 3060 $ nm via paperclip cutback structures (waveguide width of 1.8 {\textmu}m).}
\label{fig:loss3p1}
\end{figure}

Since this value is almost two orders of magnitude larger than the losses measured at $\lambda = 1550$ nm, we investigated further with Fourier-transform infrared spectroscopy (FTIR).  A SiN film 80 nm thick was deposited on a double-side-polished float zone silicon substrate and measured along with a reference silicon substrate.  The sample was oriented at 50$^{\circ}$ with respect to the illumination axis to minimize reflections.  The transmission spectrum, after baseline, reference, and water vapor corrections, is shown in Fig. S\ref{fig:loss3p1}(a).  A clear absorption signature is present near $\lambda = 3 $ {\textmu}m, which is most likely due to N-H bonds. Although the SiN film was sputtered without any hydrogenated precursors (specifically to avoid N-H bond contamination), it is possible that some hydrogen was incorporated from the residual gas background of the vaccuum chamber.  In Fig. S\ref{fig:loss3p1}(b), the absorption near $\lambda = 3 $ {\textmu}m is zoomed in to illustrate its shape.  Evidently, some absorption loss from the N-H bond is still present even at $\lambda = 3.4$ {\textmu}m, which would explain the strong impact on loss measurements at that wavelength.  We note that this sample was not annealed (which was crucial to reducing SiN losses), so it may not exhibit exactly the same characteristics as the films on the final samples.  Next, we measured loss using the DFG pump source used for mid-IR supercontinuum ($\lambda = 3.06$ {\textmu}m).  Cutback measurements of paperclip-style 1.8 {\textmu}m-wide waveguides revealed a loss of 10 dB/cm (Fig. S\ref{fig:loss3p1}(c)), confirming the stronger absorption effects for photon energies closer to the N-H resonance.

Finally, we used the FTIR measurements to estimate the limit of transparency achieved with this passivation strategy.  In Fig. S\ref{fig:loss3p1}(a), a drop in transmission is visible for $\lambda > 7.3$ {\textmu}m.  This roll-off results from a Si-N stretch resonance, an unavoidable feature of this material \cite{Verlaan2006}.  However, the technique used to realize our suspended waveguides can be readily adapted for other passivation materials with much wider transparency windows.  For example, zinc sulfide (ZnS) is transparent out to $\lambda = 13 $ {\textmu}m, at which point multi-phonon absorption causes it to become opaque \cite{bendow1977multiphonon}.  Future work will employ this or similar materials to fully unlock the transparency window of suspended AlGaAs.

\subsection{Damage threshold and physical robustness}
\label{sec:damage}

During the 1550 and 3060 nm pumped supercontinuum experiments, we kept note of the average power levels at which the input edge coupler of a waveguide was damaged.  For 1550~nm pumping, an average incident (prior to coupling) power level of 14 mW was just enough to damage some waveguide facets.  Considering the lens NA of 0.6, the 160 MHz pump repetition rate, and the pulse duration of 61 fs, this translates to a critical peak intensity of 29 GW/cm\textsuperscript{2} at the facet of the waveguide.  In earlier experiments, we also tested pumping of the waveguides with a stretched pulse of duration estimated to be 220 $\pm$ 50 fs using a lensed fiber (NA = 0.5). In this case, the maximum average power that could be tolerated increased to 70 mW. The corresponding critical peak intensity would be 31~GW/cm\textsuperscript{2}, which corresponds well with the critical intensity for the fully compressed pulse.  This suggests the failure mechanism is more dependent on peak than average power levels in the regime considered here.  Finally, we also performed the same analysis for pumping at $\lambda = $ 3060 nm.  An average incident power level of 35 mW was enough to cause facet damage.  Considering the lens NA of 0.56, the 100 MHz pump repetition rate, and the pulse duration of 81 fs, this translates to a critical peak intensity of 23 GW/cm\textsuperscript{2} at the facet of the waveguide. This places it very close to the value observed for the 1550 nm case.

The suspended membranes exhibited excellent mechanical stability during the fabrication process and subsequent experiments.  They do not collapse during liquid processing or aggressive nitrogen drying, so it is not necessary to use vapor-phase treatments at any point.  The maximum tolerable temperature (above which material defects appear) is approximately 450$^{\circ}$C, limited by thermal expansion mismatch between the materials.

%big setups fig
\begin{figure}[t]

\centering\includegraphics[width=10cm]{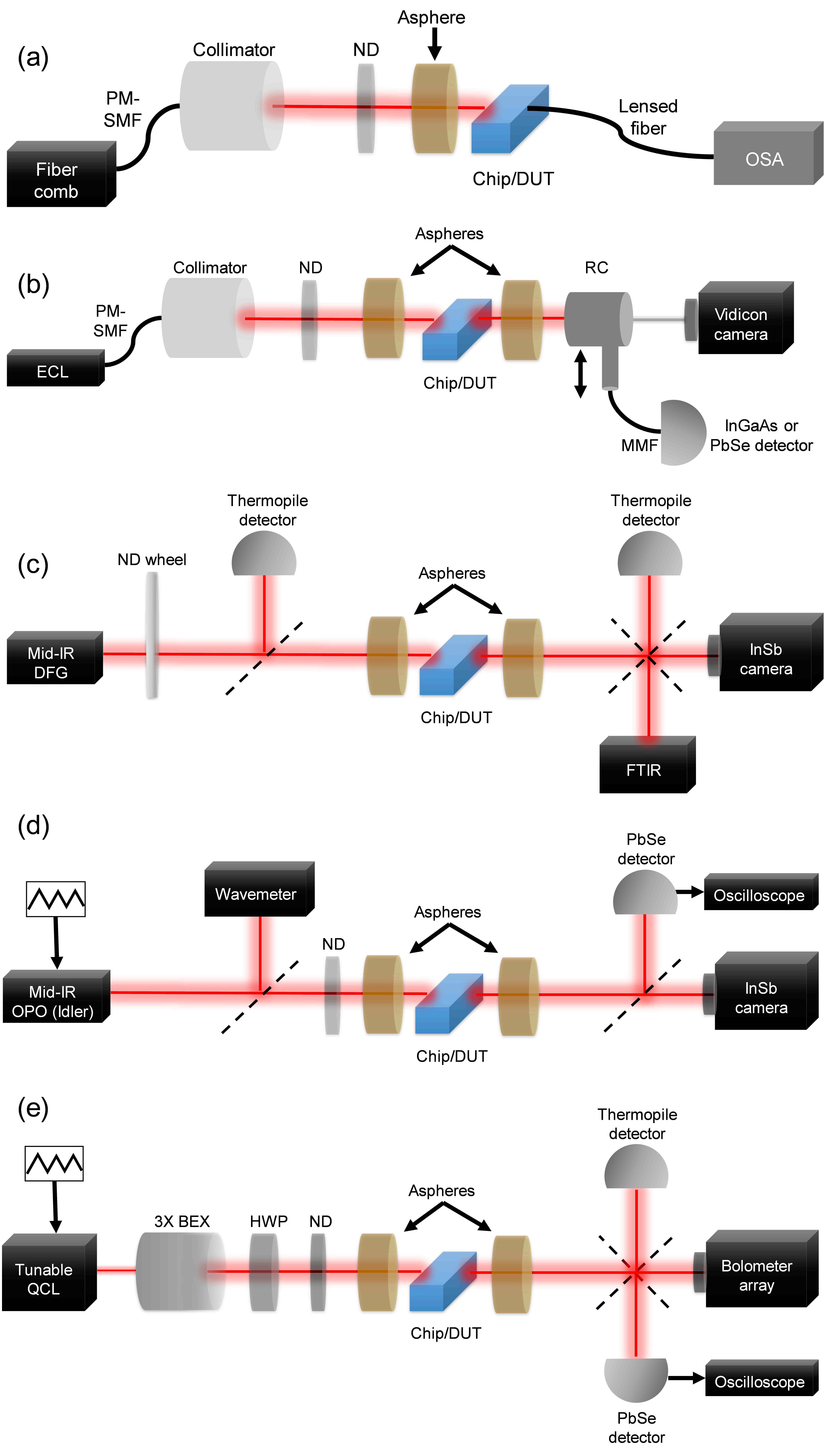}
\caption{Measurement setups used in the experiment.  (a) 1550 nm-pumped supercontinuum; (b) micro-ring resonator analysis at 1260, 1550 and 2400 nm; (c) 3060 nm-pumped supercontinuum; (d) micro-ring resonator analysis at 3600 nm; (e) micro-ring resonator analysis at 4600 nm.  PM-SMF: polarization-maintaining single-mode fiber; ND: neutral-density filter; DUT: device under test; OSA: optical spectrum analyzer; ECL: external cavity laser; RC: reflective collimator; MMF: multimode fiber; BEX: beam expander; HWP: half-wave plate.}
\label{fig:setups}
\end{figure}

\subsection{Measurement setups}
\label{sec:setups}
Here, we provide additional details on the measurement setups utilized in the experiments.  The main configurations are shown in Fig. S\ref{fig:setups}.  In all cases, the polarization of the laser light was oriented parallel to the table to excite the quasi-TE mode of the waveguides.  The setup in Fig. S\ref{fig:setups}(a) was used for the 1550-pumped supercontinuum experiments (asphere NA = 0.6, lensed fiber mode field diameter = 2.5 {\textmu}m).  Using a lensed fiber at the output allowed higher-efficiency broadband collection of light compared to a lens. The setup in Fig. S\ref{fig:setups}(b) was used for microring resonator measurements and coupling loss tests for $\lambda =$ 1260, 1550 and 2400 nm (asphere NA = 0.6).  Alignment was performed at near-IR wavelengths using the Vidicon camera, after which a reflective collimator was substituted into the beam path and used to collect the output into a 50 {\textmu}m-core multi-mode silica fiber.  The setup in Fig. S\ref{fig:setups}(c) was used for 3060 nm pumping experiments (asphere NA = 0.56). The setup in Fig. S\ref{fig:setups}(d) was used for microring resonator analysis at $\lambda =$ 3.6 {\textmu}m using the idler beam output of an optical parametric oscillator (OPO). Because of the erratic nature of wavelength tuning using etalon angles, a wavemeter was aligned to the input path to periodically check wavelength as needed.  Finally, the setup in Fig. S\ref{fig:setups}(e) was used for microring resonator analysis at $\lambda =$ 4.6 {\textmu}m.  Using laser pump current modulation at a fixed temperature, a wavelength tuning range of about 8 nm was achieved. The beam exiting the QCL was magnified 3$\times$ with a  Galilean beam expander, to better match the 6 mm clear aperture of the asphere (NA = 0.56).  The polarization was rotated to TE using a half-wave plate prior to coupling into the waveguide.

For all setups used for measuring coupling loss, the experiment was performed in four steps: (1) measure the absolute power transmission through the device under test with a power meter, (2) measure the absolute power transmission through the same path with no device in place (the light was simply coupled through both identical aspheric lenses), (3) calculate transmission as the ratio of these two quantities, and (4) factor out any relevant propagation losses (usually negligible, except for the case of $\lambda =$ 4.6 {\textmu}m).  The final loss per edge coupler is then half of the total loss.

\end{document}